\documentclass{epl}
\usepackage{epsfig}
\usepackage{amssymb}
\usepackage{amsmath}

\title{Thermal transport in granular metals}

\shorttitle{Thermal transport in granular metals}

\author{I.~S.~Beloborodov\inst{1} \and A.~V.~Lopatin\inst{1}\and
F.~W.~J.~Hekking\inst{2} \and Rosario Fazio\inst{3} \and
V.~M.~Vinokur\inst{1}}

\shortauthor{I.~S.~Beloborodov \etal}

\institute{ \inst{1} Materials Science Division, Argonne National
Laboratory, Argonne, Illinois 60439 \\
\inst{2} Laboratoire de Physique et Mod\'elisation des Milieux
Condens\'es, CNRS \& Universit\'e Joseph Fourier, BP 166, 38042
Grenoble-cedex 9, France\\
\inst{3} NEST-INFM \& Scuola Normale Superiore, I-56126 Pisa,
Italy  }


\begin{document}

\maketitle

\begin{abstract}
We study the electron thermal transport in granular metals at
large tunnel conductance between the grains, $g_T \gg 1$ and not
too low a temperature $T > g_T\delta$, where $\delta$ is the mean
energy level spacing for a single grain. Taking into account the
electron-electron interaction effects we calculate the thermal
conductivity and show that the Wiedemann-Franz law is violated for
granular metals. We find that interaction effects  suppress the
thermal conductivity  less than the electrical conductivity.
\end{abstract}

\date{\today}

The electrical conductivity $\sigma$ and the thermal conductivity
$\kappa$ of an ordinary metal are related to each other via a
universal relation~\cite{Abrikosov} known as the Wiedemann-Franz
law:
\begin{equation}
\label{law} \kappa / \sigma = L_0  T\, ,
\end{equation}
where $L_0=\pi^2 / 3 e^2 $ is the Lorentz number, $T$ is the temperature and $e$ is
the electron charge (we use units such that $k_B = \hbar =1$).  The Wiedemann-Franz
(WF) law holds as long as scattering processes are (quasi-) elastic i.e. only weakly
energy-dependent.

Equation~(\ref{law}) is a direct consequence of Fermi liquid
theory. This theory describes the properties of a weakly
disordered, interacting electron gas, provided that screening
renders the Coulomb interactions sufficiently weak and
short-ranged. Under these conditions, the low-lying excitations of
the metal are non-interacting fermionic quasi-particles that carry
both charge and energy, and as a consequence the WF law is
valid~\cite{Langer62}. In other words, no additional information
is obtained from a measurement of the thermal conductivity that is
not already present in the electrical conductivity as long as
interaction effects result only in the renormalization of electron
spectral parameters.

By now it is well known that the simultaneous action of disorder
and interactions changes the behavior of metals at low
energies~\cite{Altshuler85}. Interference of diffusively scattered
electron waves weakens the screening properties of the electron
gas, thereby leading to an energy-dependent increase of the
interaction strength. Therefore, in contrast to an ordinary
Fermi-liquid, both transport and thermodynamic quantities acquire
an additional non-trivial energy-dependence. An example is the
so-called zero-bias anomaly: close to the Fermi level Coulomb
effects suppress the tunneling density of states. Another example
is the temperature-dependent interaction correction to the Drude
conductivity.

Since the low-energy properties of disordered metals with interactions deviate from
those of an ordinary Fermi-liquid, one is, in view of the above, naturally lead to the
conclusion that the Wiedemann-Franz law (or deviations from it) may serve as an
important probe uncovering the underlying microscopic nature of disordered metallic
systems. Although this problem has been addressed by several
authors~\cite{Castellani87,Livanov91,Arfi92,Smith03,Catelani04} over the past years,
it has not been solved completely as yet. Ref.~\cite{Castellani87} finds that the WF
law holds, whereas in Refs.~\cite{Livanov91,Arfi92,Smith03,Catelani04} deviations from
the WF relation are found (although these papers do not agree on the precise form of
the deviations). We refer the reader to Ref.~\cite{Smith03,Catelani04} for a detailed
account of the present status of the field.

In this Letter we investigate this issue from a somewhat different
perspective considering thermal transport in granular metals. A
granular metal consists of small metallic grains coupled via
tunnel junctions~\cite{Efetov83}. The electron tunnelling between
grains induces randomness while the small electrical capacitance
associated with the tunnel-coupled grains induces Coulomb charging
effects. The interest in the physics of granular metals is
two-fold. First, their properties are generic for a wealth of
disordered strongly correlated systems. On the other hand, as the
interaction strength and degree of disorder can be controlled by
properly choosing the parameters of the tunnel junctions and the
grains, these systems offer a unique experimentally tunable
disordered interacting system.

Depending on the value of the dimensionless tunnel conductance
$g_T$ between the grains, we can distinguish a low conductivity
regime, corresponding to $g_T \ll 1$ and a metallic, highly
conducting regime, where $g_T \gg 1$. Electric transport in the
limit $g_T \ll 1$ has been studied extensively~\cite{Ambegaokar,
Abeles75} since the early work by Mott~\cite{Mott69}. More recent
work deals with the electronic properties of granular systems in
the metallic
regime~\cite{Beloborodov99,Efetov,Efetov02,Beloborodov03}.

In what follows we focus on the electron thermal transport in the
metallic regime, where the dimensionless tunnel conductance is
large, $g_T \gg 1$. We show that in the absence of interactions
the Wiedemann-Franz law holds and the thermal conductivity is
given by
\begin{equation}
\label{kappa0}
 \kappa_0 = L_0 \sigma _0 T,
\end{equation}
where $\sigma _{0}=2 e^{2}g_{T}a^{2-d}$ is the classical Drude
conductivity for a granular metal (including spin), $a$ is the
size of a single grain, and $d$ is the dimensionality of the
system.

We find further that interactions violate the Wiedemann-Franz law
in granular metals. Specifically, our result for the thermal
conductivity can be conveniently formulated in terms of the
quantity
\begin{equation}
\label{delta} \delta \kappa = \kappa - L_0 T\, \sigma \;\; ,
\end{equation}
that reflects deviations from the Wiedemann-Franz law. Including interaction effects,
$\delta \kappa$ in Eq.~(\ref{delta}) is given by
\begin{equation}
\label{main} \delta\kappa = \left\{
\begin{array}{lr}
 \gamma T /a
\hspace{2.2cm} d=3, &  \\
\frac{T}{3} \ln \left(\frac{g_{T} E_C }{T}\right) \hspace{1.0cm}
d=2,
\end{array}
\right.
\end{equation}
where $E_C$ is the charging energy of an isolated grain and
$\gamma \approx 0.355$ is a numerical constant. Our main
result~(\ref{main}) is valid at not too low temperatures, $T > g_T
\delta$, where $\delta$ is the mean level spacing of the grains.
At these relatively high temperatures~\cite{Efetov02}, the
electronic motion is coherent within the grains, but coherence
does not extend to scales larger than the size of a single grain.
Under these conditions, the electric conductivity $\sigma$ in the
right hand side of Eq.~(\ref{delta}) is given by the expression
obtained in Refs.~\cite{Efetov02,Beloborodov03}
\begin{equation}
\label{conductivity} \sigma = \sigma_0[\,1 - (1/2\pi d g_T)\ln(g_T
E_C/T)\,],
\end{equation}
where the electric conductivity $\sigma _{0}$ was defined below
Eq.~(\ref{kappa0}). Contrary to the electric conductivity of
granular metals, $\sigma$ given by Eq.~(\ref{conductivity}), the
thermal conductivity $\delta\kappa$ in Eq.~(\ref{main}) is
sensitive to the long-ranged character of Coulomb interaction and
therefore depends on the dimensionality of the sample, even at
temperatures $T > g_T\delta$.

One can see from Eqs.~(\ref{delta}) and (\ref{main}) that the
thermal conductivity increases with temperature in both two and
three dimensions. From Eq.~(\ref{main}) it follows that for
granular films both $\delta\kappa$ and the correction to the
electric conductivity (second term in the right hand side of
Eq.~(\ref{conductivity})) have the same logarithmic behavior such
that in $2d$ the thermal conductivity can be written as
\begin{equation}
\kappa_{2d}=\kappa_0-{{(\pi-2) }\over{ 6 }}\, T \ln\left(\frac{g_T
E_C}{T}\right), \label{kappa2D}
\end{equation}
where $\kappa_0$ is defined by Eq.~(\ref{kappa0}).

Having summarized the main results, we now turn to the
quantitative description of our model and derivation of
Eq.~(\ref{main}):  We consider a $d$-dimensional array of metallic
grains. The motion of electrons inside the grains is diffusive and
they can tunnel between grains.
 Interaction effects stem from the
long-range part of the Coulomb interaction, which is just the
classical electrostatic charging energy of the grains. Such an
array of weakly coupled metallic grains is described by the
Hamiltonian $\hat{H} = \sum _i \hat{H}_i$, where the sum is taken
over all grains in the system and
\begin{eqnarray}
\label{hamiltonian} \hat{H}_i = && \sum_{k}\xi_k a^{\dagger}_{i,k}
a_{i,k} + {\frac{{\ e^{2}}}{{\
2}}}\,\sum_{j}\,\hat{n}_{i}\,C_{ij}^{-1}\,
\hat{n}_{j} \nonumber \\
&& + \sum_{j\ne i,p,q} t_{ij}^{pq} a^{\dagger}_{i,p} a_{j,q}.
\end{eqnarray}
The first term in the right hand side of Eq.~(\ref{hamiltonian})
describes free electrons in the $i$-th isolated disordered grain,
$a^{\dagger}_{i,k} (a_{i,k})$ are the creation (annihilation)
operators for an electron in the state $k$ and $\xi_k = k^2/2m -
\mu$ with $\mu$ being the chemical potential. The second term
describes the charging energy with the capacitance matrix $C_{ij}$
and the electron number operator $\hat n_i$ in the $i$-th grain.
The last term represents the tunneling between the grains, where
$t_{ij}^{pq}$ is the tunnel matrix element corresponding to the
points of contact of $i$-th and $j$-th grains.

\begin{figure}[tbp]
\hspace{-0.5cm}
\includegraphics[width=4.0in]{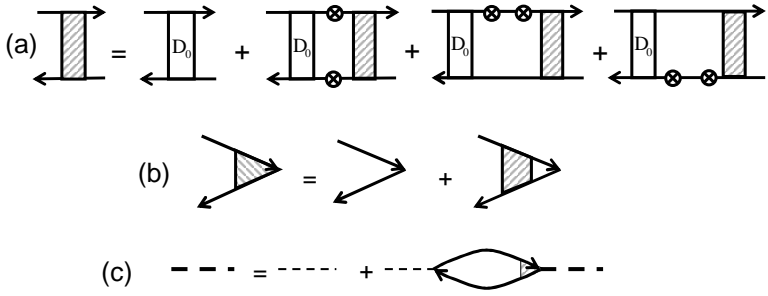}
\caption{Diagrams representing: (a) the diffusion propagator of
granular metals $D(\Omega_n, {\bf q})$ in Eq.~(\ref{diffuson}).
The tunneling vertices are described by the circles; (b)
interaction vertex renormalized by impurities and intergranular
scattering; (c) the effective screened Coulomb interaction for
granular metals $V(\Omega_n, {\bf q})$ in Eq.~\ref{Vscreen}. }
\end{figure}
The thermal conductivity of a granular metal is given by the
analytical continuation of the Matsubara thermal current-current
correlator. For granular metals the thermal current operator $\hat
J^h_{ij}$ between the $i$-th and the $j$-th grain can be obtained
as follows. The energy content of a given grain $i$ changes as a
function of time, such that
\begin{equation} d\hat{H}_i/dt = i [\hat{H},\hat{H}_i] .
\label{encons}
\end{equation}
Energy conservation requires that this energy flows to the other
grains in the system. From a straightforward calculation of the
commutator in Eq.~(\ref{encons}) we obtain $\hat J^h_{ij}$, the
amount of energy flowing per unit time from grain $i$ to a
neighboring grain $j$. It can be written as
\begin{subequations}
\begin{equation}
\label{current} \hat J^h_{ij} = \hat J^{(1)}_{ij} + \hat
J^{(2)}_{ij},
\end{equation}
where $\hat J^{(1)}_{ij}$ is the heat current in the absence of
electron-electron interaction effects
\begin{equation}
\label{J1} \hat J^{(1)}_{ij} = {i\over 2} \sum_{p,q} (\xi_{p}+\xi_{q}) \; t_{ij}^{pq}
\; a^\dagger_{i,p} a_{j,q} + h.c,
\end{equation}
while the second term, $\hat J^{(2)}_{ij}$ on the right hand side
of Eq.~(\ref{current}) appears due to Coulomb interaction effects
\begin{equation}
\label{J2} \hat J^{(2)}_{ij}= -{1\over 4} \sum_{l}
 \left [ \hat n_i C^{-1}_{il}
 \hat J_{l j}
-\hat n_j C^{-1}_{jl} \hat J_{li} \right ] + h.c.
\end{equation}
In equation~(\ref{J2}) the operator ${\hat J}_{ij}$ is the
electric current which is given by the following expression
\begin{equation}
\hat J_{ij}= i \sum_{p,q} \; t_{ij}^{pq}
 \;
a^\dagger_{i,p} a_{j,q} +h.c.
\end{equation}
\end{subequations}

For large tunnel conductance, the Matsubara thermal
current-current correlator can be analyzed by virtue of the
perturbation expansion in $1/g_T$, using the diagrammatic
technique developed in Refs.~\cite{Beloborodov99,Efetov} which we
briefly outline here. The average electron Green function has the
same form as that of a homogenously disordered metal with self
energy $\Sigma(\omega_n)=-(i/2\tau) \, {\rm sgn} \, \omega_n,$
where $\omega_n$ is a fermionic Matsubara frequency. The elastic
mean free time $\tau$ is mainly determined by the processes of
scattering by impurities inside the grain. Intergranular
scattering processes (tunneling) are included assuming that
tunneling elements are random Gaussian variables correlated
according to
\begin{equation}
\langle \, t^{p_1 p_2}_{ij}\, t^{p_3 p_4}_{ji}\rangle = t^2  \delta_{p_1 p_4}
\delta_{p_2 p_3}.
\end{equation}
Here $i$ and $j$ are the indices of the neighboring grains and $t$
is the tunnel amplitude, which is related to the dimensionless
tunnel conductance, $g_T = 2 \pi t^2/\delta ^2$. Inclusion of
intergranular scattering processes results in a small
renormalization of the mean free time $ \tau$. The diffusive
motion inside a single grain is given by the ladder diagram that
results in the diffusion propagator $ D_0^{-1} = \tau|\Omega_n|$
where $\Omega_n$ is a bosonic Matsubara frequency. The coordinate
dependence in $D_0$ is neglected since in the regime under
consideration all characteristic energies are less than the
Thouless energy. The complete diffusion propagator is given by the
ladder diagrams shown in Fig.~1a resulting in the following
expression
\begin{equation}
\label{diffuson} D^{-1}(\Omega_n, {\bf q}) = \tau(|\Omega_n|+
\varepsilon_{\bf q}\delta),
\end{equation}
where $\varepsilon _{{\bf q}}=2 g_{T}\sum_{\mathbf{a}}(1-\cos
\mathbf{qa})$ with $\mathbf{a}$ being the lattice vectors is the
function of quasi-momentum ${\bf q}$ that appears due to the
tunneling between the grains. The same ladder diagrams describe
the renormalized interaction vertex in Fig.~1b. This vertex is
used to obtain the polarization operator, that defines the
effective dynamically screened Coulomb interaction for granular
metals, Fig.~1c, 
 \begin{equation}
 \label{Vscreen}
 V(\Omega_n, {\bf q})=\left[ { { C({\bf q})} \over {e^2}} + {{ 2 \varepsilon_{
 \bf q}  }\over {|\Omega_n|+\delta \varepsilon_{\bf q} }}  \right]^{-1},
 \end{equation}
where $C(\bf q)$ is the Fourier transform of the capacitance
matrix.

\begin{figure}[tbp]
\includegraphics[width=3.0in]{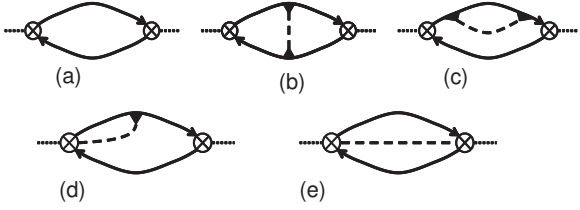}
\vspace{0.6cm} \caption{Diagrams describing the thermal
conductivity of granular metals at temperatures $T > g_T\delta$:
the diagram (a) corresponds to $\kappa_0$ in Eq.~(\ref{law}).
Diagrams (b)-(e) represent  first order corrections in $1/g_T$ to
the thermal conductivity  of granular metals due to
electron-electron interaction. The solid lines denote the
propagator of electrons, the dashed lines describe effective
screened electron-electron propagator and the black triangles
describe the elastic interaction of electrons with impurities. The
tunneling vertices are described by the circles. } \label{fig:2}
\end{figure}

In the absence of electron-electron interactions, the thermal
conductivity is given by the diagram (a) in Fig.~\ref{fig:2}. A
direct calculation of this diagram confirms that the thermal
conductivity is given by $\kappa_0$ in Eq.~(\ref{kappa0}) and thus
that the WF law holds. At temperatures $T >g_T \delta$ the first
order interaction corrections to the thermal conductivity are
given by the diagrams (b-e) in Fig.~2. We neglect the diagrams
that can be obtained from (b-e) by insertion of an extra diffusion
propagator since their contribution is small in the parameter
$g_T\delta /T$. The diagrams (b) and (c) in Fig.~2 stem from
contributions of the non-interacting part $\hat{J}_{ij}^{(1)}$ to
the thermal current-current correlator. These diagrams are
analogous to the ones considered in Ref.~\cite{Beloborodov03} for
the interaction correction $\delta \sigma$ to the conductivity of
granular metals. In fact, we find that the contribution of
diagrams (b) and (c) to $\delta \kappa$ in Eq.~(\ref{delta}) is
cancelled exactly by the corresponding term $L_0 T \delta \sigma$.
Thus, although these diagrams lead to corrections to both $\sigma$
and $\kappa$, they do not lead to a violation of the WF law.
Diagrams (d) and (e) contain the interacting part
$\hat{J}_{ij}^{(2)}$ of the thermal current. There are no
analogous diagrams for electric conductivity, hence any
non-vanishing contribution from (d) and (e) inevitably leads to a
violation of the WF law. The diagram (d) in Fig.~2 turns out to be
 zero. Thus, among the diagrams (b-e) in Fig.~2 only the diagram
(e) contributes to $\delta \kappa$ in Eq.~(\ref{delta}) giving
rise to the correction
\begin{subequations}
\begin{equation}
\delta \kappa= - {{g_T a^{2-d}}\over{2\pi T^2 }} \sum_{\bf q}
\int{{\Omega\, d\Omega}\over {\sinh^2(\Omega/2T )}} \,  {\rm Im}
\tilde V_{\bf q}(\Omega), \label{correction}
\end{equation}
where the sum is taken over the quasi-momentum ${\bf q}$ and the
potential $\tilde V_{\bf q}(\Omega)$ is given by the following
expression
\begin{equation}
\label{tildeV}
 \tilde V_{\bf q} (\Omega) = {{-i \Omega\,
[1+\cos(q_x a_x) ]\, E_C({\bf q}) }\over{ -i \Omega + 4 E_C({\bf
q}) \varepsilon_{\bf q} }}.
\end{equation}
\end{subequations}
Here $E_{C}(\mathbf{q})=e^{2}/2C(\mathbf{q})$ is the charging
energy. For the $3d$ case we simplify the right hand side of
Eq.~(\ref{tildeV}) taking the limit $E_C({\bf q }) \to \infty$.
Performing integration over $\Omega$ and summation over ${\bf q}$
in Eq.~(\ref{correction}) we obtain our final result for the
thermal conductivity (\ref{main}) in the $3d$ case, where the
numerical coefficient
$$ \gamma={{\pi a^3 }\over{ 6 }} \int {{d^3 q }\over{(2\pi)^3 }}\,
{{1+\cos(q_x a_x) }\over{\sum_{\bf a}[1-\cos({\bf q} {\bf a})]
}}\approx 0.355 .
$$
For the $2d$ case one can see that the main contribution to the
integral over ${\bf q}$ comes from the region of small
quasi-momenta $q \ll a^{-1}$ where one can use the asymptotic
expression for the capacitance matrix $C({\bf q})= a^2 q/2\pi .$
Performing the integrations in Eq.~(\ref{correction}) with
logarithmic accuracy we obtain the final result for the correction
to the thermal conductivity of $2d$ granular metals~(\ref{main}).

So far, we ignored the fact that electron-electron interactions also renormalize the
chemical potential $\mu$. Generally, this renormalization may affect the thermal
conductance:  the thermal current vertex, Eq.~(\ref{J1}), as well as the electron
Green functions depend on $\mu$. To first order in the interactions, the
renormalization of $\mu$ only leads to corrections to diagram (a) in Fig.~\ref{fig:2}.
As it can be easily shown, for this diagram the renormalization of the two heat
current vertices is exactly cancelled by the renormalization of the two electron
propagators. Therefore, the renormalization of the chemical potential by the
interactions does not affect our results in lowest order.

In this paper we only consider the electron contribution to
thermal conductivity. As it is well known, phonons will provide an
independent, additional contribution to thermal transport.
Although it is generally argued that phonons can be ignored at low
temperatures, this point needs to be considered with care as we
will now argue. Consider acoustic phonons that propagate with
velocity $c_{ph}$ in a three-dimensional impure host,
characterized by a phonon scattering length $l_{ph}$. In this
simple model, the phonon contribution to the thermal conductivity
is given by $\kappa _{ph} = T^3 l_{ph}/c_{ph}^2$. We see that
$\kappa _{ph} \sim \kappa _0$ at a temperature $T^* \sim \sqrt{g_T
c_{ph}^2/l_{ph} a}$. The electron contribution to thermal
transport dominates only for $T < T^*$. In particular, if the
phonon contribution is to be ignored in the temperature range
relevant to us, we need to impose $g_T \delta < T^*$. It is
clearly a matter of system parameters whether or not this
condition will be satisfied in a typical experiment. Avoiding the
phonon contribution in a measurement of thermal conductivity thus
implies constraints on the choice of materials as well as on
sample fabrication.

In conclusion, we have investigated the electron thermal transport in granular metals
in the limit of large tunneling conductance between the grains and temperatures $T >
g_T\delta$. We have calculated the thermal conductivity of granular metals and shown
that the Wiedemann-Franz law in granular metals is violated if electron-electron
interaction effects are taken into account. Our result can be expressed conveniently
through the Lorentz number of the granular metal, $L = \kappa /\sigma T$. In the
absence of interactions, the Lorentz number is a universal constant, $L=L_0 = \pi^2/ 3
e^2$. In the presence of interactions, we find that $L = L_0 + \delta L$, where
$\delta L = const$ for a three-dimensional granular system, whereas $\delta L \sim \ln
(g_T E_C/T)$ for a granular film. In both cases, $\delta L >0$, indicating that
interactions suppress electric conductivity more than the thermal conductivity.

Since interacting granular metals behave differently from homogeneous disordered
metals, we can not directly compare our results with the ones obtained in
Refs.~\cite{Castellani87,Arfi92,Livanov91,Smith03,Catelani04}. Nevertheless, the
structure of our result for thermal conductivity, as given by Eqs.~(\ref{delta}) and
(\ref{main}), is consistent only with that of the results of
Ref.~\cite{Smith03,Catelani04}. Specifically, we find two contributions of opposite
sign  related to interactions. One contribution is the interaction correction to
electric conductivity, multiplied by $T L_0$. The other contribution leads to a
violation of the WF law and is logarithmic in temperature for the $2d$ case, as in
Ref.~\cite{Smith03}. Of course the prefactors are different, and a complete agreement
can be expected only in the low temperature limit, $T < g_T \delta$, which is beyond
the scope of this Letter.

Experiments on thermal transport are important since they allow
for a direct observation of deviations from ordinary Fermi liquid
behavior. However, measurements of the electron thermal
conductivity of a mesoscopic conductor at low energy is still an
experimental challenge. So far, only two-dimensional electron gas
systems were studied~\cite{Molenkamp92}; more recent experiments
aimed at achieving the accuracy that is necessary to observe
deviations from the WF law~\cite{Appleyard98}. We hope that our
results that reveal deviations from Fermi liquid behavior of
granular metals will stimulate further experiments on thermal
transport.

We thank R. Ferone, A. Finkelshtein, Yu. Galperin, V. Kozub and A. Varlamov for useful
discussions. This work was supported by the U.S. Department of Energy, Office of
Science through contract No.~W-31-109-ENG-38 and by EU through contracts
RTN2-2001-00440, HPRN-CT-2002-00144. FH thanks the Materials Science Division of
Argonne National Laboratory for hospitality and acknowledges support from Institut
Universitaire de France.

\end{document}